\documentclass[smallextended]{svjour3}

\usepackage{url}
\usepackage[charsperline=0]{jlcode}
\addlitjlmacros{@constraint}{@constraint}{11}
\addlitjlmacros{@variable}{@variable}{9}
\addlitjlmacros{@objective}{@objective}{10}
\usepackage[utf8]{inputenc}

\newcommand{\revision}[1]{\textcolor{black}{#1}}

\begin{document}

\title{JuMP 1.0: Recent improvements to a modeling language for mathematical optimization}

\titlerunning{JuMP 1.0}

\author{Miles Lubin, Oscar Dowson, Joaquim Dias Garcia, Joey Huchette, Beno\^it Legat, Juan Pablo Vielma}
\authorrunning{M.~Lubin, O.~Dowson, J.D.~Garcia, J.~Huchette, B.~Legat, J.P.~Vielma}

\institute{
    Miles Lubin \at
    \email{miles.lubin@gmail.com}
    \and Oscar Dowson \at
    \email{o.dowson@gmail.com}
    \and Joaquim Dias Garcia \at
    PSR
    \and Joey Huchette \at
    Google Research
    \and Beno\^it Legat \at
    KU Leuven
    \and Juan Pablo Vielma \at
    Google Research
}

\date{Received: date / Accepted: date}

\maketitle

\begin{abstract}
JuMP is an algebraic modeling language embedded in the Julia programming language. JuMP allows users to model optimization problems of a variety of kinds, including linear programming, integer programming, conic optimization, semidefinite programming, and nonlinear programming, and handles the low-level details of communicating with solvers. After nearly 10 years in development, JuMP 1.0 was released in March, 2022. In this short communication, we highlight the improvements to JuMP from recent releases up to and including 1.0.
\end{abstract}

\section{Introduction}

JuMP is an algebraic modeling language for mathematical optimization embedded in the Julia programming language \cite{bezansonJuliaFreshApproach2017}. JuMP supports a range of common problem classes, including linear programming, integer programming, conic optimization, semidefinite programming, and nonlinear programming, and transparently handles the low-level details of communicating with the underlying solvers.  Dunning et al.~\cite{dunningJuMPModelingLanguage2017} presented JuMP as of version 0.12. Subsequently, over 90\% of JuMP's code was rewritten during the transition to a new solver abstraction layer called \textit{MathOptInterface}. 

\revision{Legat et al.~\cite{legat2021mathoptinterface} describe the design and motivation of MathOptInterface with a focus on the lower-level details that are not visible to most users of JuMP.} In this short communication, we describe key, user-visible improvements to JuMP between version 0.12 (released in February, 2016) and 1.0 (released in March, 2022). These improvements were made possible largely by the transition to MathOptInterface and other recent developments in the Julia ecosystem.

The paper is organized as a series of short sections, in which we: present a brief example demonstrating the syntax of JuMP (Section \ref{sec:examples}); 
discuss \textit{direct mode}, which reduces memory overhead and allows direct access to the low-level API of a solver (Section \ref{sec:direct-mode});
discuss how we overcome Julia's compilation latency (Section \ref{sec:latency}); update a series of benchmarks first reported in \cite{dunningJuMPModelingLanguage2017} (Section \ref{sec:benchmarks});
discuss JuMP's new attribute system (Section \ref{sec:attributes});
discuss solver callbacks (Section \ref{sec:callbacks});
and present a number of ways in which JuMP has been extended to new problem domains and applications (Section \ref{sec:extensions}). We conclude in Section \ref{sec:summary} with a brief summary of other minor---but still notable---improvements in recent releases.

\section{JuMP syntax, through example}\label{sec:examples}

To illustrate the basic usage of JuMP, the following code snippet solves a linear regression problem with additional constraints using JuMP and the nonlinear solver Ipopt \cite{wachter2006implementation}. 
\begin{lstlisting}[language = Julia]
using JuMP, Ipopt
function constrained_linear_regression(A, y)
    m, n = size(A)
    model = Model(Ipopt.Optimizer)
    @variable(model, 0 <= x[1:n] <= 1)
    @variable(model, residuals[1:m])
    @constraint(model, residuals .== A * x .- y)
    @constraint(model, sum(x) == 1)
    @objective(model, Min, sum(residuals[i]^2 for i in 1:m))
    optimize!(model)
    print(solution_summary(model))
    return value.(x)
end
A, y = rand(30, 20), rand(30)
x = constrained_linear_regression(A, y)
\end{lstlisting}
The macro syntax (lines beginning with \texttt{@}) for creating variables, adding constraints, and the setting the objective has not changed substantially since it was described by Dunning et al. \cite{dunningJuMPModelingLanguage2017}. However, the API for setting the solver, solving, and querying the results differs following the transition to MathOptInterface (between JuMP version 0.18 and 0.19). 
\section{Direct mode}\label{sec:direct-mode}

In its default usage, JuMP stores its own copy of the problem data, which enables JuMP to perform problem modification and allows the solver to be changed at any time. \revision{This design is the most typical choice for modeling languages, we presume because it simplifies the task of abstraction across solvers.}

To provide an option to reduce peak memory usage and improve performance, JuMP 1.0 includes \texttt{direct\_model}, a mode in which JuMP acts as a stateless wrapper around the solver, passing all problem data directly to the solver without any intermediate caches or transformations. \texttt{direct\_model} \revision{also} provides a safe way to interleave JuMP code with direct access to the underlying API of the solver. For example, we can call the \texttt{Highs\_scaleCol} function of HiGHS's \cite{huangfu_2018} C API as follows:
\begin{lstlisting}[language = Julia]
using JuMP, HiGHS
model = direct_model(HiGHS.Optimizer())
@variable(model, x)
#  An object that acts as a pointer to the highs problem
highs = backend(model)
#  0-indexed column of x in highs
col = HiGHS.column(highs, optimizer_index(x))
#  Call the C API
Highs_scaleCol(highs, col, 0.1)
\end{lstlisting}
\revision{Direct mode is not enabled by default because some solvers do not support the incremental modifications required to build the model in direct mode (e.g., they may be one-shot solvers that require all problem data in a single function call). Moreover, in direct mode, the user cannot change the solver associated with a model, and reformulations---particularly those associated with conic models---are disabled. In general, we suggest that users consider direct mode only if benchmarking demonstrates that model construction is a bottleneck in their application and only if they are using a solver that supports incremental modification.}

\section{Compilation latency}\label{sec:latency}

Julia was designed to be a ``fast'' language, but we find that many first-time users complain of a sluggish experience, particularly when running code from the command line or during the first invocation of a function during an interactive session. The reason is compilation latency due to Julia's Just-In-Time (JIT) compilation model.

As an example of this latency, consider the following linear program with two variables and two constraints:
\begin{lstlisting}[language = Julia]
using JuMP, HiGHS
model = Model(HiGHS.Optimizer)
@variable(model, x >= 0)
@variable(model, 0 <= y <= 3)
@objective(model, Min, 12x + 20y)
@constraint(model, c1, 6x + 8y >= 100)
@constraint(model, c2, 7x + 12y >= 120)
optimize!(model)
open("model.log", "w") do io
    print(io, solution_summary(model; verbose = true))
end
\end{lstlisting}
Saving the problem in \texttt{model.jl} and calling from the command line results in:
\begin{lstlisting}
$ time julia model.jl
15.78s user 0.48s system 100% cpu 16.173 total
\end{lstlisting}

Clearly, 16 seconds is a large overhead to pay for solving this trivial model. However, the compilation latency is independent of the problem size, and so 16 seconds of additional overhead may be tolerable for larger models that take minutes or hours to solve.

The recently matured \texttt{PackageCompiler.jl} package provides a work\-around for Julia's compilation latency by generating a custom \textit{sysimage}, a binary extension to Julia that caches compiled code. A custom image for our problem can be created as follows:
\begin{lstlisting}[language = Julia]
using PackageCompiler, Libdl
PackageCompiler.create_sysimage(
    ["JuMP", "HiGHS"],
    sysimage_path = "customimage." * Libdl.dlext,
    precompile_execution_file = "model.jl",
)
\end{lstlisting}
When Julia is run with the custom image, the run time is now 0.7 seconds instead of 16:
\begin{lstlisting}
$ time julia --sysimage customimage model.jl
0.68s user 0.22s system 153% cpu 0.587 total
\end{lstlisting}

Note that \texttt{create\_sysimage} only needs to be run once, and \revision{because the sysimage caches an intermediate representation of the code that is independent of the run-time data,} the same sysimage can be used---to a slight detriment of performance---even if we modify \texttt{model.jl} or run a different file. 

\section{Benchmarks}\label{sec:benchmarks}

In this section we update a selection of benchmarks from the previous JuMP paper \cite{dunningJuMPModelingLanguage2017}. The results are given in Table~\ref{tab:benchmarks}. \texttt{Model} is the default way of constructing a JuMP model, and, as explained in Section \ref{sec:direct-mode}, \texttt{direct\_model} is a new feature of JuMP that skips an additional cache of the problem. \texttt{Pyomo} is a popular open-source algebraic modeling language in Python \cite{hart2017pyomo}, and \texttt{GRB/C++} is the C++ interface to Gurobi \cite{gurobi} based on operator overloading.

We used Julia v1.6.2, \texttt{JuMP.jl} v1.0.0, \texttt{Gurobi.jl} v0.11.2, Gurobi v9.5.1, Python 3.8 and Pyomo v6.4.0. To remove compilation latency, we use a custom sysimage, as described in Section \ref{sec:latency}. The benchmarks were run on a 2018 Macbook Pro with 16 GB of RAM. All code to run the benchmarks is available at \url{https://github.com/jump-dev/JuMPPaperBenchmarks}.

\begin{table}[!ht]
    \centering
    \begin{tabular}{l | r | c c | c c}
        & & \multicolumn{2}{c|}{JuMP}& \\
        Model & \# Variables & \texttt{direct\_model} & \texttt{Model} & Pyomo & GRB/C++\\
        \hline
        fac-25          &    67,651 &  1 &  1 &   4 &  0 \\
        fac-50          &   520,301 &  4 &  7 &  34 &  3 \\
        fac-75          & 1,732,951 & 14 & 20 & 128 &  8 \\
        fac-100         & 4,080,601 & 37 & 50 & 338 & 20 \\
        &&& \\
        lqcp-500        &   251,501 &  2 &  1 &  24 &  3 \\
        lqcp-1000       & 1,003,001 &  7 & 11 & 107 & 10 \\
        lqcp-1500       & 2,254,501 & 17 & 23 & 234 & 23 \\
        lqcp-2000       & 4,006,001 & 32 & 50 & 417 & 40 \\
    \end{tabular}
    \caption{Time (sec.) to generate each model and pass it to the solver. Models use Gurobi and terminate after a time limit of 0.0 seconds. \revision{\texttt{direct\_model} uses the direct mode feature discussed in \ref{sec:direct-mode}, and \texttt{Model} is the default way of constructing a JuMP model}.}
    \label{tab:benchmarks}
\end{table}

Our results are broadly consistent with those reported in Dunning et al.~\cite{dunningJuMPModelingLanguage2017} and Jusevi\v{c}ius et al.~\cite{oberdieck_2021}\footnote{This is after accounting for the fact that Jusevi\v{c}ius et al.~\cite{oberdieck_2021} include compilation latency for JuMP and tested the time taken to write a JuMP model to file instead of the time taken for the solver to start performing useful work.}. JuMP is faster than Pyomo, especially on larger models, and similar in performance to the C++ interface to Gurobi. JuMP in \texttt{direct\_model} is faster than \texttt{GRB/C++} for the \texttt{lqcp} family of models, likely due to a more efficient implementation of the way in which JuMP builds quadratic expressions.

\section{Solver attributes}\label{sec:attributes}

Prior to switching to MathOptInterface, JuMP treated solver parameters non-uniformly across solvers, and supported only a fixed set of variable and constraint attributes. In JuMP 1.0, options can be bundled with a solver constructor and changed after construction using \texttt{set\_optimizer\_attribute}:
\begin{lstlisting}[language = Julia]
optimizer = optimizer_with_attributes(Ipopt.Optimizer, "max_iter" => 3)
model = Model(optimizer)
set_optimizer_attribute(model, "max_iter", 10)
\end{lstlisting}
This flexibility is particularly useful for meta-solvers (solvers that call other solvers), \revision{because it allows multiple, independent copies of a solver with different parameter settings to be created from the same input, and also because it unifies the way in which meta-solvers can pass parameters from the user through to their inner solvers}.

In addition to a number of solver-independent attributes (including starting points, verbosity, and time limits), the attribute system is extensible, allowing users to access solver-specific attributes:
\begin{lstlisting}[language = Julia]
model = Model(Gurobi.Optimizer)
@variable(model, x)
MOI.set(model, Gurobi.VariableAttribute("BranchPriority"), x, 1.0)
\end{lstlisting}

\section{Callbacks}\label{sec:callbacks}

Earlier versions of JuMP supported \revision{a range of} solver-independent callbacks, which let users write custom functions that interact with mixed-integer linear programming solvers during the solution process. \revision{Despite the promised solver inde\-pendence of the callbacks, subtle implementation differences between solvers led to different observed behavior and meant that the callbacks were in fact closely tied to the solver. This made it hard to write meta-solvers that utilized solver-independent callbacks.} In JuMP 1.0, we reduced the scope of solver-independent callbacks to a narrower subset of functionality \revision{that is more uniformly implemented by the solvers} in order to increase robustness and consistency across solvers. 

JuMP 1.0 supports three types of solver-independent callbacks: a lazy constraint callback, which allows the user to add new constraints at nodes in the branch-and-bound tree; a user-cut callback, which allows the user to add cuts which tighten the continuous relaxation at a node without removing any integer feasible solutions; and a heuristic callback, which allows the user to inject an integer feasible point. An example of a lazy constraint callback is as follows:
\begin{lstlisting}[language = Julia]
using JuMP, GLPK
model = Model()
@variable(model, 0 <= x[1:2] <= 2.5, Int)
@objective(model, Max, x[2])
function my_callback_function(cb_data)
    if callback_node_status(cb_data, model) != MOI.CALLBACK_NODE_STATUS_INTEGER
        return
    end
    x_val = callback_value.(cb_data, x)
    if x_val[2] - x_val[1] > 1 + 1e-6
        con = @build_constraint(x[2] - x[1] <= 1)
        MOI.submit(model, MOI.LazyConstraint(cb_data), con)
    elseif x_val[2] + x_val[1] > 3 + 1e-6
        con = @build_constraint(x[2] + x[1] <= 3)
        MOI.submit(model, MOI.LazyConstraint(cb_data), con)
    end
    return
end
MOI.set(model, MOI.LazyConstraintCallback(), my_callback_function)
# Would also work with CPLEX.Optimizer, Gurobi.Optimizer, etc.
set_optimizer(model, GLPK.Optimizer)
optimize!(model)
\end{lstlisting}

In addition to solver-independent callbacks, access to the \revision{low-level} API \revision{of a solver} allows the user to write solver-\textit{dependent} callbacks. For example, here we add a callback to GLPK which terminates the solve when an integer-feasible solution is detected:
\begin{lstlisting}[language = Julia]
using JuMP, GLPK
model = Model(GLPK.Optimizer)
@variable(model, 0 <= x[1:2] <= 2.5, Int)
@objective(model, Max, 1.0 * x[2])
function my_callback_function(cb_data)
    if GLPK.glp_ios_reason(cb_data.tree) == GLPK.GLP_IBINGO
        GLPK.glp_ios_terminate(cb_data.tree)
    end
    return
end
MOI.set(model, GLPK.CallbackFunction(), my_callback_function)
optimize!(model)
\end{lstlisting}

\section{Extensions}\label{sec:extensions}

A core design principle of JuMP was that it should be extensible by external packages. Therefore, JuMP exposes a number of code hooks that let users extend JuMP to new problem domains and syntax. One example is the \texttt{UnitJuMP.jl}\footnote{\url{https://github.com/trulsf/UnitJuMP.jl}} package, which adds unit information to JuMP variables and automatically scales terms in expressions to have the same units:
\begin{lstlisting}[language = Julia]
using UnitJuMP, HiGHS
model = Model(HiGHS.Optimizer)
@variable(model, x >= 0, u"m/s")
@variable(model, y >= 0, u"ft/s")
@constraint(model, x + y <= 60u"km/hr", u"km/hr")
@constraint(model, x <= 0.5y)
@objective(model, Max, x + y)
optimize!(model)
\end{lstlisting}

JuMP extensions can also combine multiple JuMP models in novel ways. For example, the \texttt{BilevelJuMP.jl} \cite{diasgarcia2022bileveljump} package lets users formulate and solve bilevel optimization problems using JuMP's syntax:
\begin{lstlisting}[language = Julia]
using JuMP, BilevelJuMP, Gurobi
model = BilevelModel(Gurobi.Optimizer, mode = BilevelJuMP.SOS1Mode())
@variable(Lower(model), x)
@variable(Upper(model), y)
@objective(Upper(model), Min, 3x + y)
@constraint(Upper(model), x <= 5)
@constraint(Upper(model), y <= 8)
@constraint(Upper(model), y >= 0)
@objective(Lower(model), Min, -x)
@constraints(Lower(model), begin
     x +  y <= 8
    4x +  y >= 8
    2x +  y <= 13
    2x - 7y <= 0
end)
optimize!(model)
\end{lstlisting}

Other popular JuMP extensions include: \texttt{InfiniteOpt.jl}, for problems with infinite-dimensional variables like continuous time and integrals \cite{pulsipher2022unifying}; \texttt{PowerModels.jl}, for solving steady-state power network optimization problems \cite{coffrin2018powermodels}; and \texttt{SDDP.jl}, for multistage stochastic programming \cite{dowson2021sddp}.

\section{Additional improvements}\label{sec:summary}

We conclude by noting additional improvements added between v0.12 and v1.0 that are minor in scope but that address longstanding feature requests.

JuMP 1.0 added the ability to delete variables and constraints, and the ability to modify the variable coefficients of linear and quadratic constraints. In addition, JuMP 1.0 provides the ability to query an irreducible inconsistent subset (IIS) of an infeasible model (if supported by the solver) and lets users access multiple primal solutions from a mixed-integer programming solver.

\revision{The number of supported solvers, as tracked by the JuMP documentation, increased from 13 to 41, and, in addition, JuMP supports any AMPL- or GAMS-compatible solver through \texttt{AmplNLWriter.jl} and \texttt{GAMS.jl}. All supported C/C++ open-source solvers are now automatically built for multiple platforms and the compiled binaries are stored online. Solver binaries are downloaded when the solver wrapper package is added by the Julia package manager. This change simplified the installation of solvers by users, greatly reducing the number of complaints of installation issues.}

The change to MathOptInterface enabled JuMP to easily support a much wider set of constraint types as well as user-defined constraint sets. Two new built-in, non-trivial constraint types are indicator and complementarity constraints, \revision{which some solvers natively support. An indicator constraint has the form $z \Rightarrow \{a^\top x \le c\}$; if the binary variable $z$ is true, then the inequality inside \texttt{\{\}} holds. A complementarity constraint has the form $f(x) \perp x_i$. If $x_i$ is at its lower bound, then $f(x) \ge 0$; if $x_i$ is at its upper bound, then $f(x) \le 0$; else $f(x) = 0$.} We also added support for generalized inequalities, in which $X \ge Y, \mathcal{S}$ is equivalent to $X - Y \in \mathcal{S}$.
\begin{lstlisting}[language = Julia]
model = Model()
@variable(model, x[1:2] >= 0)
@variable(model, z, Bin)

# Indicator constraints
@constraint(model, z => {sum(x[i] for i in 1:2) >= 1})

# Complementarity constraints
@constraint(model, complements(2x[1] + 1, x[2]))

# Generalized inequalities
X, Y = [x[1] x[2]; x[2] x[1]], [1 0; 0 1]
@constraint(model, X >= Y, PSDCone())
\end{lstlisting}

\revision{The transition to MathOptInterface also improved JuMP status reporting. Early version of JuMP relied on a free-form \texttt{Symbol} with some generic values and other solver-dependent values. Statuses are now reported using in a more unified, robust and expressive form using MathOptInterface's enums describing the status of primal and dual results and the termination status.}

Finally, we have greatly improved our documentation, which now has a wide range of how-to, reference, and tutorial materials. We recommend users get started by going to \url{https://jump.dev}.

\section*{Acknowledgements}

JuMP and MathOptInterface are open-source projects that are made possible by volunteer contributors. We thank everyone who has been part of the JuMP community during the last 10 years. Special thanks are due to the list of over 300 people who contributed code to JuMP and related projects, which is available at \url{https://jump.dev/blog/1.0.0-release/}.

\bibliographystyle{spmpsci}
\bibliography{main.bib}

\section*{Ethics declarations}

\subsection*{Funding}

O.~Dowson and B.~Legat acknowledge funding from NSF under grant OAC-1835443 for work that directly contributed to the JuMP 1.0 release.

\subsection*{Conflict of interest}

The authors declare that they have no conflict of interest.

\subsection*{Availability of data and materials}

All data analyzed during this study are publicly available. URLs are included in this published article.

\subsection*{Code availability}

All software written by the authors for this article is open source. All third-party software used in this article is open source, with the exception of Gurobi, which is commercial software available for academic use. Specific references are included in this published article.

\end{document}